# Esports and Expertise: What Competitive Gaming Can Teach Us About Mastery

Ben Boudaoud - bboudaoud@nvidia.com - NVIDIA
Josef Spjut - jspjut@nvidia.com - NVIDIA
Joohwan Kim - imjoohwankim@gmail.com - NVIDIA
Arjun Madhusudan - amadhus2@ncsu.edu - NCSU
Benjamin Watson - bwatson@ncsu.edu - NCSU

## Introduction

Historically, much research and development in human computer interaction has focused on atomic and generalizable tasks, where task completion time indicates productivity. However, the emergence of competitive games and esports reminds us of an alternative perspective on human performance in HCI: mastery of higher-level, holistic practices. Just as a world-renowned artist is rarely evaluated for their individual brush strokes, so skilled competitive gamers rarely succeed solely by completing individual mouse movements or keystrokes as quickly as possible. Instead, they optimize more task-specific skills, adeptly performing challenges deep in the learning curve for their game of choice.

Mastery or expertise refers to the acquisition of comprehensive knowledge or ability in a certain art, technique, or task. Experts typically differentiate themselves from others via high levels of performance specific to a given activity, but not necessarily through their underlying simpler and more general skills. For example, elite chess players have superior recall of chess piece configurations drawn from actual matches, but not of random configurations [Chase73]. Musical savants are superior at recalling tonal sequences following western scale structure, but not with unfamiliar sequences that violate western musical conventions. Similarly, skills in esports are most easily transferred when in-game tasks are not just mechanically, but contextually alike. Thus the perceived superiority of elite esports athletes can be better understood by considering gaming tasks holistically as opposed to dissecting these tasks into atomic actions.

Expertise is best achieved with well-known tools with high skill ceilings (i.e., tools that do not limit the individual's performance). This, in part, explains the rarity of equipment changes in traditional sports. Similarly, competitive gaming seldom changes its interfaces. Instead esports athletes spend hours mastering the human-computer interface specific to their game, given their play style. We believe that esports presents a unique opportunity to HCI researchers: to return focus to the support of task mastery. Rather than perpetually improving interfaces to support ease of learning and low-level efficiency, understanding how to enable deep expertise with well-known, high-skill ceiling interfaces should also be emphasized. By studying mastery of esports, researchers can support mastery in a broad range of other tasks.

At first glance, this suggestion may appear self-contradictory. How are we to accelerate and improve mastery in interfaces without changing them? The answer to this riddle lies in thinking beyond the interface. Computerized tasks provide unique opportunities to measure and coach performance, for example by allowing exact reproduction of challenging tasks for training purposes. By leveraging this highly controlled nature of tasks together with precise, granular and repeated measures of user performance, computers provide a unique opportunity to enable learning environments that carry users far deeper into mastery than most traditional sports coaching.

# Background

Human-computer interaction has long been driven by ease-of-use, with interfaces skeuomorphically representing well-known analog affordances, minimizing barriers to entry and accelerating acquisition of expertise early in the learning process. Examples include the pointer-based desktop interface, touch-based mobile interfaces, joysticks and steering wheels for flight and car simulators, and pressure or distance sensitive pads for drawing. In all of these cases the human-computer interface borrows from an interaction in the physical world that, at some point, was novel, requiring skilled mastery by humans.

Like many HCI researchers, our own prior work studies atomic actions, categorizing the effects of system parameters such as latency, frame rate, resolution, and display size on fundamental first-person aiming tasks. However in conducting this work, we have begun to realize that these parameters make up only a small portion of mastery of competitive gameplay. For example a static targeting task, amongst the best studied in HCI literature, rarely mimics competitive gameplay wherein targets are often other users intentionally trying to create difficult to follow motion paths. Similarly, a well-controlled and predictable targeting dynamic such as a point-and-click model does not always do a good job of mimicking the complex weapon dynamics of kick, spread, and areas of effect often found in actual competitive first-person shooter (FPS) titles. Figure 1 demonstrates some of the critical differences between static point-and-click and common FPS targeting tasks (ignoring target motion and weapon dynamics). A Fitts' style width and distance can be defined for both tasks, but the 3D nature of the first-person targeting task makes the perceived experience very different.

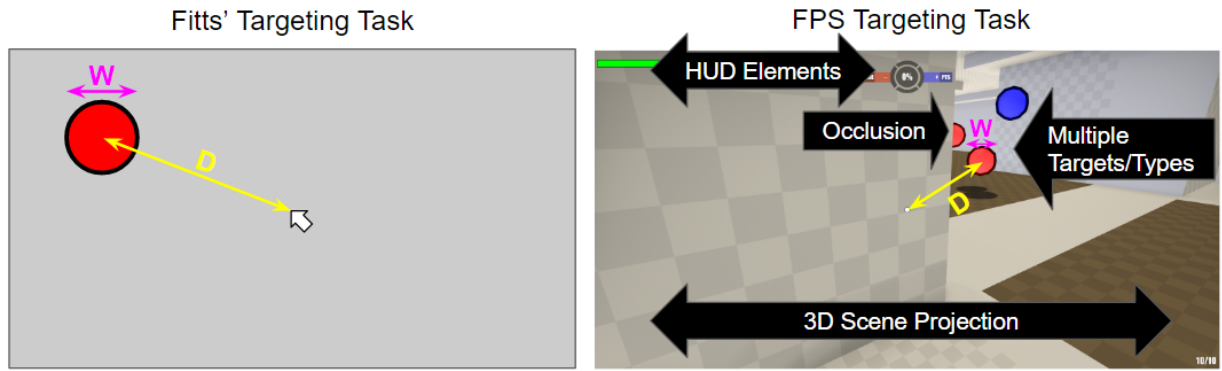

**Figure 1:** A comparison of a "pure" Fitts' law and FPS target task, both of which have been demonstrated to conform to Fitts' law. While both tasks have and respect a log(D/W) difficulty formulation, the FPS targeting task includes substantially more distractors, 3D effects, and scene dynamics.

HCI does have a tradition of research supporting mastery, dating back at least to Doug Englebart, the progenitor of much of the modern computer interface. In his "mother of all demos" in 1968, he presented interface components such as the chorded keyboard that were designed to support expertise rather than ease of learning [Eng68]. However, likely because contemporary computer interfaces were so hard to learn, research and industry focused on components that were "user friendly," such as mice and menus. For the rest of his life, Englebart bemoaned this trend, asking "would you bring a tricycle to a bicycle race?" [Eng88]. With a few exceptions, this trend in HCI has continued to today.

In focusing on mastery, Englebart also advocated for intelligence augmentation (IA) over artificial intelligence (AI) [Eng68]. Mastery implies deep human engagement in a task, while the predominant forms of AI seek to free humans from their tasks. Instead, Englebart proposed that computers help humans with their tasks — indeed, that they help them achieve mastery. In competitive gaming the use of AI is nearly always seen as anticompetitive and unfair. Alternatively, IA approaches such as console aim assist have had a warmer reception.

# Current Trends

Today, the quest for mastery is flourishing in competitive gaming and esports. The recent explosion of esports popularity and prize pools has created a renaissance of support for mastery, including a focus on higher level, game-specific tasks, computer training aids, and improved interfaces. Many expert gamers are already performing pseudoscientific studies on themselves in an attempt to gain a novel edge in performance. The broader community has demonstrated not just an interest in, but a hunger for, better-controlled scientific study of optimizing performance with their specific interfaces. This interest comes in part from the competitive nature of esports, where benefits that might be unimportant to more typical computer users can be the difference between winning and losing. Esports tasks are designed to differentiate skill levels, amplifying small differences near the limit of human mechanical and cognitive mastery (e.g., a powerful or long range weapon penalizing missed shots by enforcing a long inter-shot period). By amplifying small performance differences deep into the learning curve, esports motivate users to chase every possible performance improvement. Figure 2 provides a simplified demonstration of how the logarithmic nature of the skill-time curve can be pushed towards linear reward over time using an exponential reward-skill curve.

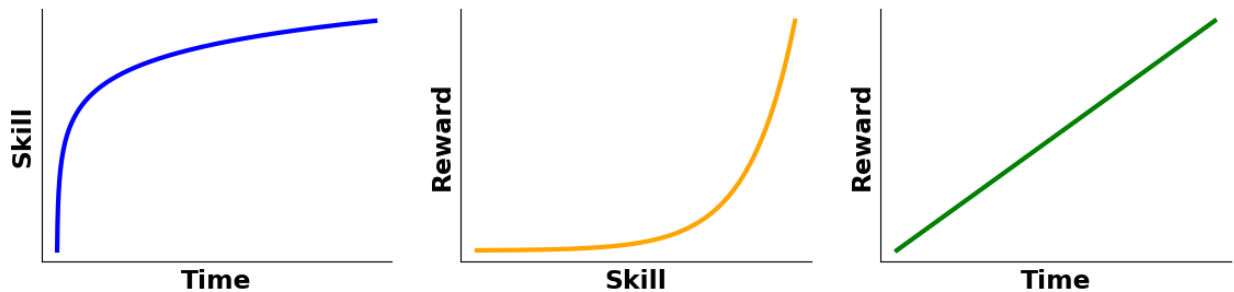

**Figure 2:** A metaphorical illustration of how esports game design tends to amplify skill differences, shaping the reward-skill curve to counteract the logarithmic nature of the skill-time (learning) curve for most applications, creating a more linear relationship in the reward-time space and driving continued improvement in user interaction.

Higher-level tasks are now an important part of competitive esports training. While initially developed only to build low-level aiming skills, aim trainer software such as KovaaK's and Aim Lab now support training under conditions more closely resembling particular games. Competitive gamers regularly scrimmage (or "scrim") to hone their skills, particularly to practice their role in the team. Examples from MOBAs (multiplayer online battle arenas) include team coordination (verbal and otherwise); team composition, in which players may take on different roles; focusing damage on particular opponents; and actions that impair the vision or movement of enemies. In games with economics, such as Counter-Strike, Starcraft, League of Legends, Dota 2 and Valorant, players may practice while better or more poorly equipped, based on economic conditions. Similarly, scrimmaging provides opportunities to train for different placement of players around maps and for unique strategic challenges. Even in single player competitive games like Starcraft, pros often play countless hours of solo matches to develop different "ramp" strategies for quickly building up resources and units to attack other players, exhaustively searching for slightly better ways to save time in building up their in-game advantage. The practice of "theorycrafting" has become popular, wherein players create spreadsheets and run monte-carlo simulation of various potential encounters, particularly to determine which high level strategies will give them an advantage.

New esports training tools are emerging that are tightly integrated with games themselves. Several games offer detailed match replay files and/or short video recaps ("killcams") following being defeated, allowing players to analyze their performance to improve actions in the next encounter. Platforms like Mobalytics [Mobalytics22] offer users higher-level insights into their performance by longitudinally collecting data from game servers, tracking these statistics over time as a player improves. Many other web tools provide matchup analytics for character picks, and allow players to practice the character drafting meta-game.

The quest for mastery has also driven demand for innovation in esports interfaces — without fundamentally changing interaction paradigms. For example, vendors have begun supporting customization of mouse weight and shape, and have also increased mouse reporting rates to up to 8 kHz. New displays can refresh at 360 or even 500 Hz. The performance-oriented technical specifications of such input and output devices often meet or exceed those of specialized devices used in human performance research, and are well beyond the needs of typical computer users.

In at least one case, interfaces supporting esports mastery have diverged qualitatively from interfaces supporting productivity, rather than just exceeding their needs. HCI research has long observed that pointer acceleration improves task completion time and throughput [Cas08]. (Pointer acceleration, or "nonlinear CD gain", makes cursors move farther when the mouse moves the same distance more quickly). Yet in FPS games, top players almost always turn off pointer acceleration, since it requires them to repeat not just a precise displacement, but also a certain velocity profile to perform a given action again. Even more interestingly, some more experimental FPS gamers are beginning to adopt pointer acceleration, but using speed to movement transforms that are quite different from the defaults adopted by Windows or other windowed operating systems.

With all of this support for expertise, esports is a unique opportunity to study the acquisition of task and interface mastery more generally [Cam18]. First, the data and training context is unparalleled. Other computer applications rarely integrate training for higher-level tasks, and largely ignore usage histories. Conventional sports science does an admirable job of gathering many statistics on and off the field, particularly with modern tracking technologies, but pales in comparison to the granularity and pervasiveness of esports data capture. Rich tools for analysis paired with fine-grained, longitudinal data describing competition in highly precise and memory abundant computer systems make up the unique value proposition of esports. Second, the range of user skill and variety of tasks are exceptional. The skill level of the esports user base ranges from novice computer users to professional gamers who have honed their skills for years. In traditional sports science, only elite athletes have detailed performance data tracked over time. In esports, every player does. Additionally, esports tasks are often broad and deliberately unconstrained, so data from just one title can often provide insight into a wide range of interactions.

# Future Directions

We predict that more holistic, application-specific performance models will become increasingly relevant in HCI, particularly in highly demanding and competitive mastery-oriented settings where professional coaching or individualized advice are becoming the status quo. Rather than reject this more complete but specific knowledge as "lacking broader impact" or "unlikely to generalize", we advocate the study of the techniques and tools being used by esports athletes to attain mastery, so that we may begin helping a broader set of users develop expertise with their interfaces.

Improved tools can accelerate the development of interface mastery, both in esports and other domains. For example, a standard for describing real, in-game situations would enable esports athletes to drill for specific mechanics in which they require more expertise, and might allow them to compare their actions to experts'. Alternatively, in some cases, non-interactive simulation might improve performance, helping to find near optimal choices in a broad user decision tree. Finally, while usage trace data is often collected today, more could be done to instrument the user. Such human-centric data might help us better understand how people react to challenges and successes in real time, unlocking the next level of performance through optimization on the human side of the human-computer interaction loop. To accurately affiliate human reactions to digital events, computer and human trace data must then be synchronized, a non-trivial task on its own.

Current performance-enhancing improvements to the esports interface will continue, which should have value outside of esports. Mice and displays will continue to update more frequently and with reduced delays. Researchers will attempt to learn which characteristics of improved input and output devices are most beneficial, and which displayed information enhances performance most. As with many modern workers, esports athletes are often formed into teams, requiring rapid and effective communication and creating demand for better technologies supporting collaboration.

As an aside, many have argued that augmented and virtual reality (AR/VR) will be an important part of competitive gameplay's future. It certainly is not considered mainstream in today's esports community. One possible reason for this is that AR/VR interfaces and interactions do not yet support the long-tailed development of mastery, as mice, keyboards and game controllers do. To change this, AR/VR interfaces must support prolonged use (over multiple continuous hours), reducing nausea in particular. When these interfaces eventually do improve support for development of expertise, the physicality and "naturalness" of their interactions may make them less interesting to HCI researchers studying mastery of more conventional interfaces. AR/VR experiences typically strive to reproduce real world interfaces (as a component of "presence") and in the limit, cease to behave any differently.

Which non-esports applications might benefit from new understanding, training tools and interfaces supporting expertise? Mastery is most valued in tasks that:

- *Have a long learning curve*. Simpler tasks are easier to master, giving such mastery less value. This implies that valuable mastery requires a complex task.
- *Reward continued learning.* Skill acquired deep in the learning curve has little value, if it is not rewarded. High skill rewards drive growth towards mastery.
- *Create an intrinsic desire to improve*. Without this motivation skill tends to plateau, sating extrinsic motivators. In the late stages of mastery intrinsic value becomes necessary.

We believe that digital applications such as art, coding, computer aided design and media/content creation all have these characteristics, and could benefit by borrowing techniques supporting mastery from esports. Other applications might include teleoperation and collaboration. On the other hand, typical productivity applications such as word processing and spreadsheet tools rarely require or reward continued learning and development years into their use. Of course, all of these conjectures should be examined in future research.

Supporting expertise will be an important part of future HCI research. By aiding users in their pursuit of interaction mastery, technology can leverage human adaptation to better augment their abilities. In esports, expertise is highly evolved, valued and supported, with competition driving continuous improvement. By studying the tools and techniques esports athletes use to master their interfaces, much can be learned about how to support expertise in a broader range of applied interactions. To this end, we advocate study of the often game- and task-specific techniques used by professional gamers to acquire expertise, as both an invaluable means of understanding interface mastery more broadly, and of building a brighter future for high skill HCI.

# Citations

# Author Bios

## Ben Boudaoud

Ben Boudaoud is a research engineer at NVIDIA working at the interface of human interaction and practical system design. His research interests include energy efficient embedded systems, novel interaction modalities, and high performance HCI.

## Josef Spjut

Josef Spjut is a Research Scientist at NVIDIA working on esports, graphics, and player performance. His research contributed to the RT Core hardware in Turing and newer GPUs and the NVIDIA Reflex esports platform. A lifelong gamer, he is most excited by how game interfaces connect people to each other.

## Joohwan Kim

Joohwan Kim is a research manager leading Nvidia's Human Performance and Experience research group based in Santa Clara, California. Joohwan's current interests are in understanding and improving viewer experience of various types of displays, especially regarding esports.

## Arjun Madhusudan

Arjun Madhusudan is a Ph.D. student at North Carolina State University working on esports and player performance studies. Arjun is also actively involved in an esports experience studies with NVIDIA. He received his MS from NC State as well, and has been continuing this field of research since.

## Benjamin Watson

Benjamin Watson is Associate Professor of Computer Science at North Carolina State University. His interdisciplinary Visual Experience Lab focuses on the engineering of visual meaning, and works in the fields of graphics, visualization, interaction and user experience. Watson co-chaired the ACM Interactive 3D Graphics and Games (I3D) 2006 conference.